\documentclass[12pt]{iopart}
\usepackage{iopams,graphicx,subfigure,color,mathptmx,url,cite,harvard,hyperref}

\newcommand{\openone}{\leavevmode\hbox{\small1\normalsize\kern-.33em1}}

\begin{document}

\title{Physical mechanisms underpinning the vacuum permittivity}

\author{Gerd Leuchs$^{1,2,*}$, Margaret Hawton$^{3}$ and Luis~L~S\'{a}nchez-Soto$^{1,4}$}

\address{$^{1}$ Max-Planck-Institut f\"{u}r die Physik des Lichts, Staudtstra\ss e 2, 91058~Erlangen, Germany}

\address{$^{2}$ Institute of Applied Physics, Russian Academy of Sciences, 603950 Nizhny Novgorod, Russia}

\address{$^{3}$ Department of Physics, Lakehead University, Thunder Bay, Ontario P7B 5E1, Canada}

\address{$^{4}$ Departamento de \'{O}ptica, Facultad de F\'{\i}sica, Universidad Complutense, 28040~Madrid, Spain}

\ead{gerd.leuchs@mpl.mpg.de}

\date{\today}

\begin{abstract}
Debate about the emptiness of the space goes back to the prehistory of science and is epitomized by the Aristotelian \emph{horror vacui}, which can be seen as the precursor of the ether, whose modern version is the dynamical quantum vacuum. Here, we change our view to \emph{gaudium vacui} and discuss how the vacuum fluctuations fix the value of the permittivity $\varepsilon_{0}$ and permeability $\mu_{0}$.
\end{abstract}

\eqnobysec


\section{Introduction}

In textbooks of electromagnetism, it is often assumed that the vacuum permittivity $\varepsilon_{0}$ (and permeability $\mu_{0}$) are merely measurement-system constants. In this vein, both are not considered as fundamental physical properties, but rather artifacts of the SI system, which disappear in Gaussian units. However, this simplistic view ignores that, irrespective of the method of allocating a value to $\varepsilon_{0}$ (and $\mu_{0}$), they just express the prediction of Maxwell's equations that, in free space, electromagnetic waves propagate at the velocity of light.
 
From a classical perspective, this propagation warrants the existence of an ether, an all-pervading medium composed of a subtle substratum. This is a powerful explanatory concept that goes back to the prehistory of science and helped unify our understanding of the physical world for centuries~\cite{Whittaker:2012va}. Maxwell himself invoked a structured vacuum to motivate his displacement current leading to the prediction of electromagnetic waves~\cite{Maxwell:2011vc}.  

It is a standard mantra that the ether was abandoned largely because of Einstein’s special relativity, which contradicts an absolute frame of reference. Nonetheless, Einstein’s relationship with the ether was complex and changed over time~\cite{Wilczek:2008aa}. Despite its very negative connotations, the notion of ether nicely captures the way most physicists actually think about the vacuum~\cite{Laughlin:2005vo}.

In quantum electrodynamics (QED) the vacuum shows up as a modern relativistic ether~\cite{Dirac:1951vd}, although it is not called that way because it is ``taboo"~\cite{Laughlin:2005vo}. This quantum vacuum is a dynamical object, containing the seeds of multiple virtual processes~\cite{Borchers:1963aa,Sciama:1991aa,Milonni:1994vs}. Several effects manifest themselves when the vacuum is perturbed in specific ways: vacuum fluctuations lead to shifts in the atomic energy levels~\cite{Lamb:1947aa}, changes in the boundary conditions produce particles~\cite{Moore:1970aa}, and accelerated motion~\cite{Unruh:1976aa} and gravitation~\cite{Hawking:1975aa} can create thermal radiation.

The concept of zero-point energy arose before the development of the quantum formalism~\cite{Boyer:1985tl}. However, in quantum theory zero-point energy rests upon a much firmer foundation than was possible classically.  Observable phenomena, such as the Casimir effect~\cite{Milton:2001aa}, strongly suggest that the vacuum electromagnetic field and its zero-point energy are real physical entities~\cite{Milonni:1994vs}. 

QED envisages vacuum fluctuations as particle-antiparticle pairs that appear spontaneously, violating the conservation of energy according to the Heisenberg uncertainty principle. A careful discussion of the nature of these fluctuations can be found in~\cite{Mainland:2020wp}.  These pairs determine the value of $\varepsilon_{0}$ and $\mu_{0}$: a photon will feel the presence of those pairs much the same it feels polarizable matter in a dielectric. This idea can be traced back to Furry and Oppenheimer~\cite{Furry:1934aa}, Weisskopf and Pauli~\cite{Pauli:1934aa,Weisskopf:1936aa}, Dicke~\cite{Dicke:1957aa}, and Heitler~\cite{Heitler:1954vl} who mulled over the prospect of treating the vacuum as a medium with electric and magnetic polarizability. 

If the value of  $\varepsilon_{0}$ is determined by the structure of the vacuum, it should be possible to calculate it by examining the (polarizing) interaction of photons introduced into the vacuum as test particles~\cite{Mainland:2020wp}.  These ideas have been recently readressed~\cite{Leuchs:2010aa,Leuchs:2013aa,Urban:2013aa,Mainland:2017aa,Mainland:2018aa,Mainland:2019aa,Leuchs:2019aa} to calculate \textit{ab initio}  $\varepsilon_{0}$~by using methods similar to those employed to determine the permittivity in a dielectric.  Interestingly, the possibility that a charged pair can form an atomic bound state (which can thus be well approximated by an oscillator) was discussed by Ruark~\cite{Ruark:1945aa} and further elaborated by Wheeler~\cite{Wheeler:1946aa}.  In this paper, we elaborate on these ideas and claim that the value of $\varepsilon_{0}$ can definitely be estimated from first principles.

\begin{figure}[t]
\centering 
\includegraphics[width=.65 \columnwidth]{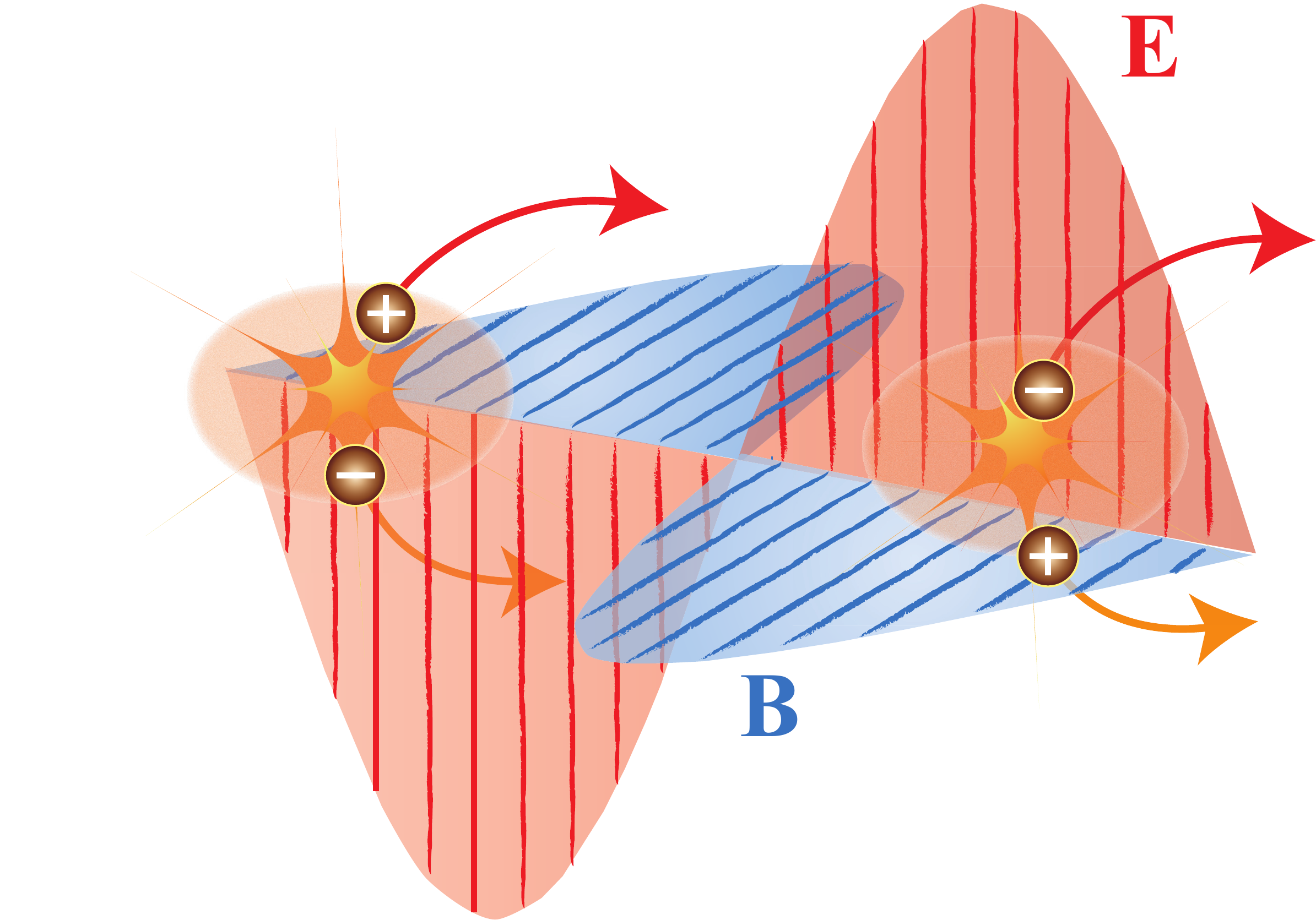}
\caption{Cartoon view of the particle-antiparticle pairs continually created in the vacuum.}  
\label{fig:vac}
\end{figure} 

\section{A simple dielectric model of vacuum polarization}
\label{sec:dielmodvac}

The modern view~\cite{Paraoanu:2015tg} interprets that particle–antiparticle pairs are continually being created in a  vacuum filled with the vacuum electromagnetic field. They live for a brief period and then annihilate one another. The lifetime of such a virtual particle pair is governed by its energy through the energy–time uncertainty principle~\cite{Hilgewoord:1996ds}
\begin{equation}
\label{eq:uncET}
\Delta \mathcal{E} \, \Delta \tau \gtrsim \hbar \, .
\end{equation}
The creation of this virtual pair requires surplus energy of at least $2 mc^{2}$, where $m$ is the mass of each partner. Energy conservation must be violated by $\Delta \mathcal{E} \gtrsim 2 m c^{2}$.  Equation~(\ref{eq:uncET}) implies that the violation is not detectable in a period shorter than $\hbar / (2mc^{2})$, so virtual particles can survive roughly that time. Since nothing can move faster than light, such a virtual pair must remain within a distance $d = \hbar/(2 mc)$; that is, a distance of order the Compton wavelength $\lambda_{\mathrm{C}} = \hbar/(mc)$. This also demonstrates that heavy pairs require a larger $\Delta \mathcal{E}$ and thus their effect is concentrated at smaller distances. 

Charged electron-positron  pairs behave much like the bound charges of atoms in a polarizable medium.  We thus assume that the physical properties of the vacuum are governed by those virtual pairs reacting to external fields just like any ordinary material but with the permittivity and permeability values $\varepsilon_{0}$ and $\mu_{0}$.  

The dipole moment induced in the electron-positron bound state can be estimated using a harmonic-oscillator model with a spring constant given by the energy gap $\hbar \omega= 2 m c^2$~\cite{Leuchs:2010aa,Leuchs:2013aa}.  We can compute the corresponding displacement $x$ by assuming the quasi-static limit of the oscillator, for which $m \omega^{2} x = e E$, where $E$ is the electric field. The resulting dipole moment is thus
\begin{equation}
\wp = e x = \frac{e^{2} \hbar^{2}}{2 m^3 c^4} E \, .
\end{equation}
Note that two equal harmonically bound masses $m$ correspond to an oscillator with reduced mass $m/2$. As a result, the dipole moment density turns out to be
\begin{equation}
P_{0} = \frac{\wp}{\lambda_{\mathrm{C}}^{3}} = \frac{e^{2}}{2 \hbar c} E \, .
\end{equation}
The quantity multiplying $E$ plays the role of an effective vacuum permittivity. Interestingly, since the mass drops out,  different types of elementary particles having the same electric charge contribute equally to the vacuum polarizability irrespective of their mass. Therefore, we can write 
\begin{equation}
\varepsilon_{0} = f \frac{1}{2 \hbar c} 
\sum_{\mathfrak{s}}^{\mathrm{e.\, p.}} q_{\mathfrak{s}}^{2}  \, , 
\label{eq:e0} 
\end{equation} 
where $f$ is just a correction factor (of order unity) that accounts for finer details and the sum is over all possible leptons with charge $q_{\mathfrak{s}}$.

One might wonder about a possible frequency dependence of the vacuum polarization as a result of the resonances at the rest mass energies.   The conservation of momentum prohibits the excitation of a virtual pair to a real pair in free space with a plane wave. Far away from resonance, the process is allowed because of the quantum uncertainty of the momentum.  In contradistinction, a converging electromagnetic dipole wave may excite real pairs in the vacuum~\cite{Narozhny:2004uu}. 

Equation~\ref{eq:e0}, with only the contribution of electron-positron pairs, gives about $18\%$ of the established value of $\varepsilon_{0}$. One could thus rightly argue that heavier particle pairs might dominate~\cite{Hajdukovic:2010tm}.  It has also been suggested that  instead of a single type of particle involved, there is a Gaussian distribution of probabilities of the vacuum energy fluctuations and, consequently, a whole range of particle pairs are produced, with the center of mass averaged to anywhere in between~\cite{Margan:2017uv}. This consideration will not alter the result in equation~\ref{eq:e0}.

\section{Vacuum polarization in QED}
\label{sec:vacpol}
 
\begin{figure}[t]
\centering 
\includegraphics[width=.60 \columnwidth]{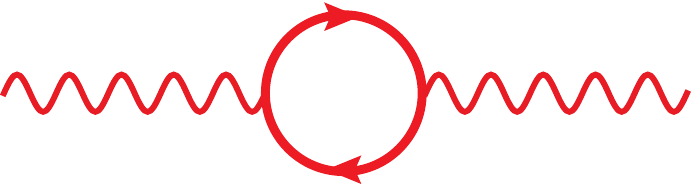}
\caption{Vacuum polarization in the one-loop approximation. The wavy lines represent an electromagnetic  field, while a vertex represents the interaction of the field with the fermions, which are represented by the internal lines. The resulting polarization is maximal for a free electromagnetic field for which $\omega =|\bi{k}|c$.} 
\label{fig:Feynman}
\end{figure}

The virtual pairs discussed qualitatively in the previous section can be nicely depicted in terms of the time-honored Feynmann diagrams. Figure~\ref{fig:Feynman} is such a representation of vacuum polarization in the one-loop approximation.  By making use of the standard machinery of Feynmann diagrams~\cite{Peskin:2018qv}, one can show that, at lowest order in $\alpha$, they induce the following susceptibility~\cite{Prokopec:2004vh} 
\begin{equation}
\fl
\chi_{e} (k^{2}, \Lambda) = 8 \pi \alpha  \int_{0}^{1} \rmd x \, x (1-x) 
\int^{\Lambda} \frac{d^{3}p}{(2 \pi)^{3}} \left [ 
\bi{p}^{2} + (mc/\hbar)^{2}+ x (1-x) k^{2} \right ]^{-3/2} \, ,
\end{equation}
where 
\begin{equation}
\alpha = \frac{1}{4\pi \varepsilon_{0}}\frac{e^{2}}{\hbar c} 
\end{equation}
 is the fine structure constant, with value $\alpha^{-1} = 137.035999084(21)$.  Note that the linear response of the vacuum, as represented by this susceptibility, must be Lorentz invariant, so in reciprocal space the susceptibility of vacuum must be a function of $k^{2} = \omega ^{2}/c^{2}-\bi{k}^{2}$.  The condition $k^{2}=0$, describing a freely propagating photon, is referred to as on-shellness: a real on-shell photon  verifies then $\omega^{2}=\bi{k}^{2}c^{2}$.

The integral over $\bi{p}$ represents the contribution from a photon of wave vector $\bi{k}$ exciting an electron with momentum $\bi{p} + \hbar x \bi{k}$ and a positron with momentm $- \bi{p} + \hbar (1-x) \bi{k}$. This process conserves the three-momentum $\bi{p}$, but not the energy. As discussed before, individual pairs with very high $|\bi{p}|$ do not contribute much because they are too ephemerals to polarize much. However, there are so many states with large momentum that their net contribution diverges: the cutoff $\Lambda$ is introduced precisely to avoid that problem. If we integrate over momenta and expands in powers of $1/\Lambda$ we get
\begin{equation}
\fl
\label{eq:chie1}
\chi_{e} (k^{2}, \Lambda) = \frac{4 \alpha}{\pi}  \int_{0}^{1} \rmd x \, 
x (1-x) \left[ \ln \left ( \frac{2 \hbar \Lambda}{mc}\right ) - 1 -
\frac{1}{2} \ln \left ( 1 + x(1-x) \frac{\hbar k^{2}}{m^{2}c^{2}} \right ) \right ] \, .
\end{equation}
We see that the susceptibility diverges logarithmically in the limit $\Lambda \rightarrow \infty$. This leads to a physically unreasonable result: the photon self-mass becomes infinite~\cite{Bogoliubov:1959aa}.  

Given the $\alpha$ in the numerator, equation~(\ref{eq:chie1}) is a statement about the product $\varepsilon_{0} \chi_{e} (k^{2}, \Lambda) $, not about the separate factors. Notice that in standard QED $\chi_{e} (0, \Lambda \rightarrow \infty) = 1$ is a constant and $e^{2}$ is renormalized instead. 

Since observations are made near $k^{2}=0$, it is costumary to take the susceptibility relative to its on-shell value,  
\begin{eqnarray} 
\chi_{e} (k^{2}) & \equiv & \lim_{\Lambda \rightarrow \infty} 
[ \chi_{e} (k^{2}, \Lambda) - \chi_{e} (0, \Lambda)] \nonumber \\
& = &  - \frac{2 \alpha}{\pi} 
\int_{0}^{1} \rmd x \, x (1-x) \left[  1  + x(1-x) \frac{\hbar k^{2}}{m^{2}c^{2}} \right ] \, , 
\label{DeltaPi} 
\end{eqnarray}
as the relevant quantity, which is independent of the cutoff. This is an archetypal example of a regularization in the theory. 

The remaining integral can be readily performed, leading to a cumbersome analytical expression~\cite{Itzykson:1980aa}. However, in the interesting limit $\hbar^{2}\left\vert \bi{k}^{2}\right\vert \gg m^{2}c^{2}$ we get the simple expression 
\begin{equation} 
\label{eq:chi2}
\chi_{e} (k^{2}) = -  \frac{\alpha}{3 \pi} 
\ln \left ( \frac{\hbar k^{2}}{A m^{2}c^{2}} \right ) \, ,   
\end{equation} 
where $A = \exp(5/3)$. 

As in a standard dielectric, this susceptibility contributes to the permittivity as  
\begin{equation}
\varepsilon_{0}(k^{2})= \varepsilon_{0}[1 + \chi_{e}(k^{2})] \, .
\label{eq:epsilon} 
\end{equation} 
The dependence of the permittivity $\varepsilon_{0}(k^{2})$ on the energy scale $k^{2}$ is known as running. Note that $\varepsilon_{0}$ corresponds to $\varepsilon_{0}(k^{2}=0)$: QED can calculate the running part via $\chi_{e}(k^{2})$, but $\varepsilon_{0}$ has to be determined experimentally. 

The standard linear relations in classical electromagnetism 
\begin{eqnarray} 
\bi{D}(k) =\varepsilon_{0}( k^{2}) \, \bi{E}(k) \, ,  \qquad  
\bi{H}(k) = c^{2} \varepsilon_{0}(k^{2}) \, \bi{B}(k) \, . 
\label{VacuumDH} 
\end{eqnarray} 
are maintained, but because of the running, $\varepsilon (k^{2}) \le \varepsilon_{0}$~\cite{Leuchs:2019aa}.

Of course, electrons and positrons are not the only kinds of charged particles. To obtain the susceptibility contributed by other kinds of spin 1/2 particles, we simply replace $m$ and $e$ in the previous expressions with the corresponding values.  Charged particles with spin zero also entail replacing the factor of $x(1 - x)$ in the integral (\ref{eq:chie1}) by $(1-2x^{2})/8$~\cite{Prokopec:2004vh}.  

In the fine structure constant $\alpha$, we hold $e$ constant and incorporate the $k$-dependence into $\varepsilon_{0}(k^{2})$.  Since $\varepsilon_{0}(k^{2})^{-1}$ contains all powers of $e^{2}$, it includes summation over all numbers of pairs. When restricted to an energy scale $\mathcal{E}_{\mathrm{max}}$, the sum is over all fermions of mass less than $\mathcal{E}_{\mathrm{max}}/c^2$~\cite{Eidelman:1995aa,Hogan:2000aa,Hoecker:2011aa}. Considering $e^{2}/\varepsilon_{0}( k^{2}) $ is in most ways equivalent to running of the square of effective charge in conventional QED, but the physical interpretation is different. In a dielectric it is possible to have $\varepsilon_{0}<0,$ but $e_{\mathrm{eff}}^{2}<0$ makes no physical sense.

The dielectric properties of vacuum differ from those of a material medium in two important ways: $\ln (k^{2})$ dependence replaces the usual $\omega $ dependence and Lorentz invariance requires that $\varepsilon_{0}( k^{2}) \mu_{0}( k^{2}) =1/c^{2}$. The speed $c$ is a universal constant  whereas the coupling constant $ \alpha(k^{2} )$ runs. On the photon mass shell  $k^{2}=0$, so a free photon always sees $\varepsilon_{0} $ and there is no running. 

Linear response theory~\cite{Nussenzveig:2012wx} suggests continuation of the function $\chi_{e} (k^{2})$ in the complex $k$ plane, so as to obtain a Kramers-Kronig  dispersion relation linking the real and imaginary parts of this susceptibility. A lengthy calculation shows that~\cite{Itzykson:1980aa}
\begin{equation}
\label{eq:impi}
\mathrm{Im}\, \widehat{\chi}_{e} (k^{2}) = 
\frac{\alpha^{2}}{3\pi}  e^{2}
\left ( 1 - \frac{4 m^{2}c^{2}}{\hbar^{2}k^{2}} \right )^{1/2}  
\left ( 1 + \frac{2 m^{2}c^{2}}{\hbar^{2}k^{2}} \right ) \, .
\end{equation}
This gives the absorptive part, independently of any regularizing cutoff, but this absorption only happens when $\hbar k > 2 m c$, which corresponds to the process of pair creation in an electric field~\cite{Sauter:1931aa,Schwinger:1951aa}.

We conclude by claiming that the simple back-of-the-envelope calculation sketched in section~\ref{sec:dielmodvac} is altogether consistent with QED. Actually, the loop in figure 1 can be thought of as a single polarizable atom with center-of-mass momentum $\hbar k$. If, for simplicity, we set $k = 0$, the computation of the Feynmann diagram involves integrals of the form $\int d^{4}q \, [q^{2} + mc/\hbar  ]^{-2}$, which entails an exponential decay $\exp [- (mc/\hbar) |\bi{x}|]$ in real space. Therefore, the ``radius" of such a virtual atom is of order $\hbar/(mc)$. All in all, this suggests that the virtual pairs can be modelled as oscillating dipoles with frequency $mc^{2}/\hbar$ and volume of order $[h/(mc)]^{3}$. This is effectively a cutoff, as it has been used in some early QED calculations.

Indeed, at large $k^{2}$ and to second order in perturbation theory, equation~(\ref{eq:chi2}) gives
\begin{equation} 
\varepsilon_{0} (k^{2}) \simeq  \varepsilon_{0} -  
\frac{1}{12\pi^2 \hbar c}  \sum_{\mathfrak{s}}^{\mathrm{e. \, p.}}    
q_{\mathfrak{s}}^{2}\;   \ln \left( \frac{\hbar^{2}k^{2}}
{m_{\mathfrak{s}}^{2}c^{2}}\right) \, ,
\end{equation} 
where we have explicitly included summation over all possible pairs. It is known that at high-momentum (or energy) scale, the coupling constant $\alpha (k)$ in QED becomes infinity~\cite{Gell-Mann:1954aa}. In physical terms, charge screening can  make the  ``renormalized" charge to vanish. This is often referred to as triviality~\cite{Bogoliubov:1959aa}. If $\Lambda_{\mathrm{L}}$ is the value of that momentum [which is usually called the Landau pole~\cite{Abrikosov:1954aa}], then~\cite{Leuchs:2017aa}  
\begin{equation} 
\varepsilon_{0} = \frac{1}{12\pi^2 \hbar c}  
\sum_{\mathfrak{s}}^{\mathrm{e. \, p.}}  
q_{\mathfrak{s}}^{2}\;  \ln \left(
\frac{\hbar^{2}\Lambda_{\mathrm{L}}^{2}}{m_{\mathfrak{s}}^{2}c^{2}}\right) \, .
\end{equation} 
If we compare with equation~(\ref{eq:e0}), it turns out that they are identical,  provided the fudge factor $f$ is identified with 
\begin{equation} 
\label{eq:7} f = \frac{1}{12 \pi^{2}} 
\frac{\displaystyle\sum_{\mathfrak{s}}^{\mathrm{e. \ p.}}
q_{\mathfrak{s}}^2 \; \ln  \left(
\frac{\hbar^{2}\Lambda_{\mathrm{L}}^{2}}{m_{\mathfrak{s}}^{2} c^{2}} \right )}
{\displaystyle \sum_{\mathfrak{s}}^{\mathrm{e. \ p.}}
q_{\mathfrak{s}}^2} \, . 
\end{equation} 
The masses $m_{\mathfrak{s}}$ differ by a factor $10^{6}$, but this factor is diminished by the logarithmic function. As a result the logarithmic term is almost constant for large enough cutoff $\Lambda_{\mathrm{L}}$ and, to some approximation, can be taken out of the sum. For the standard model $\sum q_{\mathfrak{s}}^{2}/e^{2}=9$ and with two additional charged Higgs particles of $mc^2 \simeq  5\times 10^{11}$~eV, $f \simeq 1$ and all of $\varepsilon_{0}$ is vacuum polarization if $\ln(\hbar \Lambda_{\mathrm{L}}/c)\simeq 26$.

\section{Physical interpretation of the running} 
\label{sec:GWmod}

\begin{figure}[t]
\centering 
\includegraphics[height=7cm]{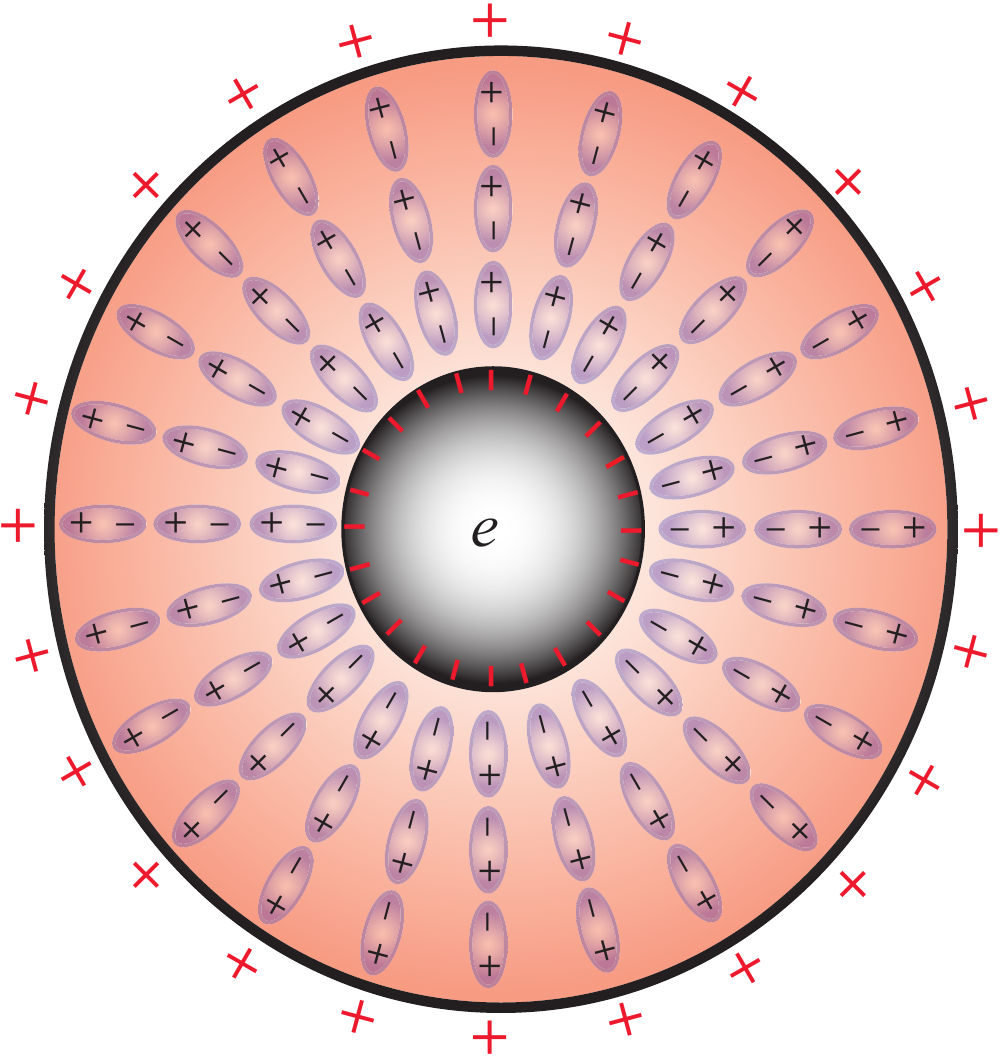}
\caption{Simple model to illustrate the vacuum polarization.} 
\label{fig:GWdia}
\end{figure} 

We try now to gain further insights into the physical mechanisms behind vacuum polarization. We discuss a simple model~\cite{Gottfried:1986aa} that considers the bare charge $e_0$ as filling a small sphere of radius $a$, with $a \ll \lambda_{\mathrm{C}}$. If this distribution is surrounded by a spherical shell of inner radius filled with a dielectric medium of permittivity $\varepsilon $, as sketched in figure~\ref{fig:GWdia}, an induced charge appears on the inner surface of the shell at radius $a$ and an equal charge of opposite sign appears at the outer surface of the dielectric medium. However, when the charge is measured  \emph{within} the medium (e.g., via a Gaussian surface) it will appear to be reduced. Although a bit trivial, this familiar example should caution us about the subtleties of defining \emph{charge} when a medium is involved. In particular, the distinction between \emph{bare} charge (that describes an isolated system) and \emph{dressed} charge (that takes into account screening in the medium) must be stressed. 

In contradistinction with an ordinary dielectric medium, one has to take into account that the vacuum extends to infinity and any observer lies inside the medium. To understand the \emph{dressing} now, we must in addition realize that the vacuum permittivity depends on the distance $r$ to the charge. This occurs  because at $r$ only those virtual pairs having Compton wavelengths $\lambda_{\mathrm{C}} \lesssim r$ contribute. 

We can  interpret this model from a QED viewpoint. To this end, we define the relationship between $ \bi{D}$ and $\bi{E}$ at charge separations $r \gg \lambda_{\mathrm{C}} $ where the coupling strength can be measured; that is,  $\varepsilon_{0}( r \gg \lambda_{\mathrm{C}}) \equiv \varepsilon_{0}$ at large distances or, equivalently, small momenta, where the vacuum is maximally polarized. With polarization included the dielectric permittivity is $\varepsilon_{0}$ at the physical scale. In consequence,   $\chi_{e} (k^{2})$ is the reduction in polarization. 

Since $\varepsilon_{0}( k^{2}) $ is not a constant, the exact relationship between $\bi{D}$ and $\bi{E}$ is nonlocal in $r$-space. We will therefore look at the problem in momentum space, where we have an expression for the permittivity. We examine this screeing for the simplest example of a static charge, say $+e.$ In the Coulomb gauge $\bi{E} (\bi{k}^2) = - \bi{k} \Phi (\bi{k}^2) $ and $\omega =0$. Since now $k^{2}=-\bi{k}^{2}$, then the Coulomb potential reads
\begin{equation} 
\Phi ( \bi{k}^{2} )= \frac{e} 
{\bi{k}^{2}\varepsilon_{0}( \bi{k}^{2})} \, .
\label{CoulombDielectric}
\end{equation} 
To get the corresponding expression in $\bi{r}$-space, we take the inverse Fourier transform 
\begin{equation}
\Phi ( \bi{r}) =\int \frac{d^{3}\bi{k}}{( 2\pi )^{3}} 
\frac{e}{\bi{k}^{2}\varepsilon_{0}( \bi{k}^{2})} \exp(
i\bi{k}\cdot \bi{r}) \, , 
\label{Coulomb_r} 
\end{equation}%
For electron-positron pairs the magnitude of $\chi_{e}(k^{2})$ is very much less that $1$ so that $\varepsilon_{0}^{-1}(k^{2})$ is approximately $ [1 - \chi_{e}(k^{2})]/\varepsilon_{0}$ and we get~\cite{Peskin:2018qv}
\begin{equation} 
\Phi  ( \bi{r} ) \simeq
\frac{e}{4\pi \varepsilon _{0}r} \times  
\left \{   \begin{array}{ll} 
\displaystyle 1 +\frac{2\alpha }{3\pi }  \ln \left( \lambda_{\mathrm{C}}/{r} \right)
-\gamma - \frac{5}{6}  & \qquad r \ll \lambda_{\mathrm{C}} , \\ & \\
\displaystyle 1+\frac{\alpha }{4\sqrt{\pi}} \frac{e^{-2 r/\lambda_{\mathrm{C}}}}{\left(r/ \lambda_{\mathrm{C}} \right)^{3/2}}  & \qquad r \gg \lambda_{\mathrm{C}} ,
\end{array} \right . 
\label{approxCoulomb} 
\end{equation}
where $\gamma =0.577$ is Euler's constant. The radiative correction to the Coulomb potential is called the Uehling potential~\cite{Uehling:1935mc}. The screening appears in a crystal-clear manner: the dielectric constant $\varepsilon_{0}$ decreases with increasing $\bi{k}^{2}$ or decreasing $r$~\cite{Landau:1973aa}. For $r<\lambda_{\mathrm{C}}$ the Coulomb interaction becomes stronger.

\section{Concluding remarks}

The vacuum permittivity has so far been a purely experimental number.  We have worked out a simple dielectric model to point at the intimate relationship between the properties of the quantum vacuum and the constants in Maxwell’s equations.  From this picture, the vacuum can be understood as an effective medium. We hope that with all these arguments, the misconception that $\varepsilon_{0}$ is just an adjusted measurement-system constant will be dismissed from physical courses and textbooks.

\ack 
GL acknowledges financial support by the Ministry of Education and Science of the Russian Federation megagrant No. 14.W03.31.0032. LLSS acknowledges financial support from  Spanish MINECO (Grant PGC2018-099183-B-I00). 

\newpage 


\end{document}